\documentclass[11pt,a4paper]{article} 
\usepackage{a4wide}
\usepackage{amsmath}
\usepackage{graphicx}
\usepackage{makecell}
\usepackage{verbatim}
\usepackage{multirow}
\usepackage{subfig}
\usepackage{bm}
\usepackage{braket}
\usepackage{mathtools}
\usepackage{amssymb}
\usepackage{amscd}
\usepackage{latexsym}
\usepackage{slashed}
\usepackage[normalem]{ulem}
\usepackage{verbatim}
\usepackage{rotating}
\usepackage{diagbox}
\usepackage{cancel}
\usepackage[utf8]{inputenc}
\usepackage{array}
\usepackage{epsfig,amsfonts,amsthm}
\usepackage{float}
\usepackage{slashed}
\newlength{\absize}
\usepackage{verbatim}
\usepackage{cite}
\numberwithin{equation}{section}

\begin{document}

\thispagestyle{empty}
\renewcommand{\thefootnote}{\fnsymbol{footnote}}
\newpage\normalsize
\pagestyle{plain}
\setlength{\baselineskip}{4ex}\par
\setcounter{footnote}{0}
\renewcommand{\thefootnote}{\arabic{footnote}}
\renewcommand{\title}[1]{
\begin{center}
\LARGE #1
\end{center}\par}
\renewcommand{\author}[1]{
\vspace{2ex}
{\Large
\begin{center}
 \setlength{\baselineskip}{3ex} #1 \par
\end{center}}}
\renewcommand{\thanks}[1]{\footnote{#1}}
\renewcommand{\abstract}[1]{
\vspace{2ex}
\normalsize
\begin{center}
\centerline{\bf Abstract}\par
\vspace{2ex}
\parbox{\absize}{#1\setlength{\baselineskip}{3.5ex}\par}
\end{center}}

\vspace*{4mm}

\title{\textbf{A $D_5$ Model of Quarks with Explicit CP Violation} }

\author{Dong-Ping Fu$^{a}$\footnote{E-mail: fudongp@mail2.sysu.edu.cn}
and
Michihisa Takeuchi$^{a}$\footnote{E-mail: takeuchi@mail.sysu.edu.cn}}

\begin{center}
$^{a}$School of Physics and Astronomy, Sun Yat-sen University, 519082 Zhuhai, China 
\end{center}

\vfill

\abstract{The hierarchy of fermion masses and mixings lacks a self-consistent theoretical explanation at the dynamical or symmetry level, and the origin of CP violation remains unclear. These two issues have long been major challenges in particle physics. In this paper, we investigate the quark sector of a four-Higgs-doublet model with an exact $D_5$ symmetry. Within the framework of explicit CP violation, we consider the most general real vacuum, and perform a numerical fit of the quark mass spectrum and the CKM matrix using a Differential Evolution. We demonstrate that the model achieves an excellent overall goodness of fit with $\chi^2 = 6.97$, and the Jarlskog invariant $J_{\mathrm{CP}} = 3.144 \times 10^{-5}$ agrees with the experimental value within $0.19\sigma$. This indicates that the new source of CP violation provided by the model is consistent to generate the observed CP violation in the quark sector. Moreover, the hierarchies inherent in the vacuum expectation values and the quark Yukawa couplings naturally explain the observed hierarchy of the quark masses and mixing parameters. }
\vspace*{30mm} \setcounter{footnote}{0} \vfill
\newpage
\renewcommand{\thefootnote}{\arabic{footnote}}

\setcounter{page}{1}
\section{Introduction}
\label{Sec:intro}
The Standard Model (SM) of particle physics is one of the most outstanding achievements of modern physics, but due to many unresolved problems, it is unlikely to be the ultimate theory. First, the origin of CP violation in the SM remains unclear, indicating the need for additional sources of CP violation beyond the SM. Experimentally, the fermion masses span many orders of magnitude from the first to the third generation, and the CKM matrix elements also exhibit a clear hierarchical pattern; these common features constitute the ``hierarchy'' of fermion masses and mixing~\cite{Kobayashi:1973fv,Cabibbo:1963yz,Gell-Mann:1964ewy,Glashow:1970gm,SLAC-SP-017:1974ind,Georgi:1974sy,E288:1977xhf}.
However, the origin of this hierarchy lacks a self-consistent theoretical explanation at the dynamical or symmetry level within the SM, which has long been a major puzzle in physics.
The problems of fermion masses, mixing, and CP violation are collectively referred to as the three central problems in flavor physics.

The SM itself cannot solve these puzzles.
Among various approaches, multi-Higgs-doublet models (MHDMs) with an extended Higgs sector have been one of the most actively investigated directions, as they can provide additional sources of CP violation~\cite{Branco:1985pf,Wu:1994ja,Branco:2005em,Nishi:2006tg,Inoue:2014nva,Barradas-Guevara:2015rea,Grzadkowski:2016szj,Emmanuel-Costa:2016vej,Okada:2020brs,Kuncinas:2023ycz,Miro:2024zka}.
Meanwhile, discrete symmetries, which include both non‑Abelian and Abelian types, are often introduced to reduce the number of free parameters in MHDMs, and can also provide a natural mechanism for explaining many flavor puzzles. In this context, many studies have sought to establish a connection between fermion masses and mixing by imposing flavor symmetries~\cite{Branco:1996bq,Frampton:1999hk,Altarelli:2005yx,Hagedorn:2006ir,Botella:2009pq,Ferreira:2010ir,Ishimori:2010au,Machado:2010uc,Keus:2013hya,Cogollo:2016dsd,Petcov:2018snn,King:2020qaj,Yao:2020qyy,Babu:2023oih,Chauhan:2023faf,Abe:2023ilq,Ding:2024pix,Rocha:2024twm,Rocha:2025ade}. This line of thought implies that at some ultra-high energy scale (e.g., the GUT or Planck scale) where the SM is to be embedded into a more fundamental theory, there may exist certain flavor symmetries. The spontaneous breaking of these symmetries at that scale constrains the matrix form of the Yukawa couplings, which after renormalization group evolution down to the TeV scale lead to specific textures and thus naturally yield the observed quark mass and mixing patterns.
Such a mechanism can provide a self-consistent theoretical origin for the hierarchy of quark masses and mixing parameters.

Motivated by this, in this work we systematically study the quark sector of a four-Higgs-doublet model with an exact $D_5$ symmetry ($D_5$~4HDM). In previous work, we first constructed the scalar potential of this model and performed a complete analysis of its neutral vacuum structure, presenting a complete list of all possible real and complex vacuum configurations, and identifying among them the complex vacua that can lead to spontaneous CP violation (with C-N-3a being the most general one), along with the corresponding constraints on the potential parameters and the positive-definiteness conditions of the Hessian \cite{Fu:2026agd}. Subsequently, we extended these vacuum configurations to the Yukawa sector of quarks and systematically investigated all possible representation assignments of the quark fields under $D_5$. We found that only when the three generations of quark fields are assigned as combinations of doublet and singlet representations can one avoid massless quarks and obtain a non-block-diagonal CKM matrix. However, in such models, due to the strong constraints imposed by the $D_5$ symmetry, all vacua that admit spontaneous CP violation (SCPV) force the up- and down-type quark mass-squared matrices to be simultaneously block-diagonalized by the same unitary transformation, thereby leading to a block-diagonal CKM matrix, and thus yielding vanishing mixing angles and CP violation \cite{Fu:2026phe}. This conclusion rules out the possibility of explaining experimental observations within the framework of SCPV, and also points the way for further exploration in the framework of explicit CP violation. On this basis, in the present work we further study the case of explicit CP violation, aiming to test whether the $D_5$~4HDM can successfully fit the quark mass spectrum and mixing parameters while maintaining a non-block-diagonal CKM matrix, and can provide new sources for the observed CP violation, thereby offering a natural symmetry-based explanation for the hierarchy of quark masses and mixings.

The remainder of this paper is organized as follows.  In Section~\ref{Sec:D5quark}, we assign specific $D_5$ charges to the quark fields, and derive the corresponding quark mass matrices under the most general real vacuum R-N-4a, and present the conditions for obtaining non-vanishing quark masses and a non-block-diagonal CKM matrix.In Section~\ref{Sec:quarkfit}, we perform a numerical fit. At the scale $\mu = 1\,\mathrm{TeV}$, using the 2024 PDG experimental data, we adopt a differential evolution algorithm to fit the quark masses and the $\mathrm{CKM}$ mixing matrix, analyze the best fit results, and further provide a natural explanation for the origin of CP violation and the hierarchy of quark masses and mixing. Section~\ref{Sec:Conclusion} summarizes our conclusions. The appendices provide further details on the \(D_5\)-symmetric scalar potential, the \(D_5\) tensor-product rules, the unitary mixing matrices at the best-fit point, and the removal of unphysical phases from the Yukawa couplings.

\section{Quark sector with $D_5$ symmetry}
\label{Sec:D5quark}
In a general four-Higgs-doublet model (4HDM), the Yukawa couplings in the quark sector can be written as a sum over all four Higgs doublets, and the Yukawa Lagrangian can be expressed as:
\begin{equation}
-\mathcal{L}_Y = \bar{q}_L \left( \sum_{i=1}^{4} \Gamma_i \phi_i \right) d'_R + \bar{q}_L \left( \sum_{i=1}^{4} \Delta_i \tilde{\phi}_i \right) u'_R +\text{h.c.},
\end{equation}
where $q_L = (u'_L, d'_L)^T$ is a column vector in the three-dimensional generation space, whose three components are the three generations of left-handed SU(2) quark doublet fields.
The SU(2) singlet $u'_R$ ($d'_R$) denotes a right-handed 
up-type (down-type) quark field, and it is a vector in the three-dimensional generation space.
The $\tilde{\phi}_i = i\sigma_2 \phi_i^*$ is the conjugate representation of the Higgs doublet $\phi_i$, and $\Delta_i$ and $\Gamma_i$ are the $3\times 3$ Yukawa coupling matrices. After spontaneous electroweak symmetry breaking, the Higgs fields acquire vacuum expectation values (VEVs) $\langle \phi_i \rangle = v_i / \sqrt{2}$, inducing the physical mass matrices for the down-type and up-type quarks:
\begin{equation}
\begin{split}
 M_d = \frac{1}{\sqrt{2}} \sum_i \Gamma_i v_i, \qquad M_u = \frac{1}{\sqrt{2}}\sum_i \Delta_i v_i^*.
    \end{split}
\end{equation}
These matrices are generally non-diagonal. We define $U_\alpha$ ($\alpha = d_L, d_R, u_L, u_R$) as the matrices that transform the quarks into the mass basis:
\begin{equation}
\begin{split}
  & \bar d'_L=\bar d_LU^\dagger_{d_L}, \qquad d'_R=U_{d_R}d_R,\\&
   \bar u'_L=\bar u_L U^\dagger_{u_L},
   \qquad u'_R=U_{u_R}u_R \ .
\end{split}
\end{equation}
The CKM matrix for quark mixing is given by
\begin{equation}
\label{equ:CKM}
  V_{CKM}=U^\dagger_{u_L}U_{d_L} \ .
\end{equation}
The mass matrices $M_d$ ($M_u$) are diagonalized by the unitary matrices $U^\dagger_{d_L}$ and $U_{d_R}$ ($U^\dagger_{u_L}$ and $U_{u_R}$). We obtain the diagonalized mass matrices for down-type and up-type quarks as follows:
\begin{equation}
\label{Dmd}
\begin{split}
&\mathcal{M}_d=\mathrm{diag}(m_d,m_s,m_b)=U^\dagger_{d_L} M_d U_{d_R}=\frac{1}{\sqrt{2}}U^\dagger_{d_L}\Bigl[\sum_i v_i\Gamma_i\Bigr]U_{d_R}, \\&
\mathcal{M}_u=\mathrm{diag}(m_u,m_c,m_t)=U^\dagger_{u_L} M_u U_{u_R}=\frac{1}{\sqrt{2}}U^\dagger_{u_L}\Bigl[\sum_i v_i\Delta_i\Bigr]U_{u_R}.
\end{split}
\end{equation}
After diagonalization, the mass-squared matrices are:
\begin{equation}
\label{Dmd2}
\begin{split}
&\mathcal{M}^2_d={\rm diag}(m^2_d,m^2_s,m^2_b)=U^\dagger_{d_L}M_d M^\dagger_d U_{d_L},\\& \mathcal{M}^2_u={\rm diag}(m^2_u,m^2_c,m^2_t)=U^\dagger_{u_L}M_uM^\dagger_u U_{u_L} \ .
   \end{split}
\end{equation}
\subsection{R-N-4a model under explicit CP violation}
In $D_5$ 4HDM, the transformations of four scalar fields under $D_5$ symmetry are given in Ref.~\cite{Fu:2026agd} as follows:
\begin{equation}
\label{Phi1234}
\begin{pmatrix}
\phi _1    \\
\phi _2    \\
\end{pmatrix}\sim \mathbf{2},\quad \quad 
\begin{pmatrix}
\phi _3    \\
\phi _4    \\
\end{pmatrix}\sim \mathbf{2}'.
\end{equation}

In Ref.~\cite{Fu:2026phe}, we systematically studied all possible representation assignments of the three generations of quark fields under the $D_5$ group, and found that only when the quark fields are assigned as combinations of doublet and singlet representations can one avoid massless quarks and obtain a non-block-diagonal CKM matrix. Among such viable representation assignments, all possible quark transformation schemes can be classified into two basic types, namely Case A and Case B in Ref.~\cite{Fu:2026phe}. In the numerical fitting study of the present work, we analyzed these two cases separately and found that only Case B can yield a quark mass spectrum and mixing parameters consistent with experimental data while maintaining a non-block-diagonal CKM matrix. Therefore, in this paper we adopt the Case B quark transformation scheme, whose $D_5$ representation assignments are given as follows:
\begin{equation}
\label{qdu22'2'}
\begin{split}
&\begin{pmatrix}
\bar q^1_{L}\\
\bar q^2_{L}    
\end{pmatrix} \sim \mathbf{2},\qquad 
\bar q^3_{L}\sim \mathbf{1},\\
&\begin{pmatrix}
d'_{1R}    \\
d'_{2R}    \\
\end{pmatrix},\ 
\begin{pmatrix}
u'_{1R}    \\
u'_{2R}    \\
\end{pmatrix} \sim \mathbf{2}',\qquad 
 d'_{3R},\  u'_{3R}\sim \mathbf{1}.
\end{split}
\end{equation}

Then, according to the rules for the tensor product of the $D_5$ group~\cite{Fu:2026agd}, the Yukawa Lagrangian of the quark sector
singlet under the $D_5$ is obtained as:
\begin{subequations}
\begin{equation}
\begin{aligned}
\mathcal{-L}_{Y_d}=\;&y_{1}^{d}(\bar q^1_{L}{\phi}_1 d'_{2R} + \bar q^2_{L}{\phi}_2 d'_{1R})+y_{2}^{d}(\bar q^1_{L}{\phi}_2 + \bar q^2_{L}{\phi} _1)d'_{3R}\\&+y_{3}^{d}(\bar q^2_{L}{\phi}_4 d'_{2R} + \bar q^1_{L}{\phi}_3 d'_{1R})+ y_{4}^{d} \bar q^3_{L}({\phi}_3 d'_{2R} + {\phi}_4 d'_{1R}) + \text{h.c.},
\end{aligned}
\end{equation}
\begin{equation}
\begin{aligned}
\mathcal{-L}_{Y_u}=\;&y_{1}^{u}(\bar q^1_{L}\tilde{\phi}_2 u'_{2R} + \bar q^2_{L}\tilde{\phi}_1 u'_{1R})+y_{2}^{u}(\bar q^1_{L}\tilde{\phi}_1 + \bar q^2_{L}{\phi}_2)u'_{3R} \\&+y_{3}^{u}(\bar q^2_{L}\tilde{\phi}_3 u'_{2R} + \bar q^1_{L}\tilde{\phi}_4 u'_{1R})+ y_{4}^{u} \bar q^3_{L}(\tilde{\phi}_4 u'_{2R} + \tilde{\phi}_3 u'_{1R}) + \text{h.c.},
\end{aligned}
\end{equation} 
\end{subequations}
where $y_{i}^{d}$ and $y_{i}^{u}$ ($i=1,2,3,4$) are the independent complex Yukawa coupling coefficients of the down-type and up-type quarks, respectively.
Ref.~\cite{Fu:2026agd} has presented the complete neutral vacuum structure of the $D_5$ 4HDMs within the framework of CP conservation in the scalar potential, including both real and complex vacua.
On this basis, since the Yukawa coupling coefficients in the quark sector considered in this work are allowed to have arbitrary complex phases, the model exhibits explicit $\mathrm{CP}$ violation.
After spontaneous electroweak symmetry breaking, we consider the most general real vacuum among them, R-N-4a, i.e., $(v_1, v_2, v_3, v_4)$, whose complete minimization conditions are given in Ref.~\cite{Fu:2026agd}.
In this vacuum, the following mass matrices for the down-type and up-type quarks are obtained:
\begin{equation}
\label{equ:quark_model_d}
M_{d}=\frac{1}{\sqrt{2}}\begin{pmatrix}
 y_{3}^{d}v_3&  y_{1}^{d}v_1&y_{2}^{d}v_2\\
y_{1}^{d}v_2&y_{3}^{d}v_4& y_{2}^{d}v_1\\
y^{d}_{4}v_4& y^{d}_{4}v_3& 0
\end{pmatrix},
\end{equation}
\begin{equation}
\label{equ:quark_model_u}
M_{u}=\frac{1}{\sqrt{2}}\begin{pmatrix}
 y_{3}^{u}v_4&  y_{1}^{u}v_2&y_{2}^{u}v_1\\
 y_{1}^{u}v_1&y_{3}^{u}v_3& y_{2}^{u}v_2\\
y_{4}^{u}v_3& y_{4}^{u}v_4& 0
\end{pmatrix}.
\end{equation}
Then, the mass squared matrices of quarks, $M_d M_d^\dagger$ and $M_u M_u^\dagger$, respectively, as follows:
\begin{equation}
\label{MdMd}
\begin{aligned}
&M_d M_d^\dagger = \frac{1}{2}
\begin{pmatrix}
|y_1|^2 v_1^2 + |y_2|^2 v_2^2 + |y_3|^2 v_3^2
& |y_2|^2 v_1 v_2 + y_1 y_3^* v_1 v_4 + y_3 y_1^* v_2 v_3
& y_4^* (y_1 v_1 + y_3 v_4) v_3 \\
|y_2|^2 v_1 v_2 + y_3 y_1^* v_1 v_4 + y_1 y_3^* v_2 v_3
& |y_2|^2 v_1^2 + |y_1|^2 v_2^2 + |y_3|^2 v_4^2
& y_4^* (y_1 v_2 + y_3 v_3) v_4 \\
y_4 (y_1^* v_1 + y_3^* v_4) v_3
& y_4 (y_1^* v_2 + y_3^* v_3) v_4
& |y_4|^2 (v_3^2 + v_4^2)
\end{pmatrix},
\end{aligned}
\end{equation}
\begin{equation}
\label{MuMu}
\begin{aligned}
&M_u M_u^\dagger = \frac{1}{2}
\begin{pmatrix}
 |y_{1}|^2 v_2^2 + |y_{2}|^2 v_1^2+|y_{3}|^2 v_4^2  & |y_{2}|^2 v_1 v_2+y_{3} y_{1}^* v_4 v_1 + y_{1} y_{3}^* v_2 v_3  &
y_{4}^* \left(  y_{1} v_2+y_{3} v_3  \right)v_4 \\ |y_{2}|^2 v_1 v_2 + y_{1} y_{3}^* v_1 v_4 + y_{3} y_{1}^* v_3 v_2 & |y_{1}|^2 v_1^2 + |y_{2}|^2 v_2^2 + |y_{3}|^2 v_3^2  &
y_{4}^*  \left( y_{1} v_1 + y_{3} v_4 \right)v_3 \\
y_4  \left( y_3^* v_3 + y_1^* v_2 \right)v_4 & y_{4} \left( y_{1}^* v_1 + y_{3}^* v_4 \right) v_3&
|y_{4}|^2 (v_3^2 + v_4^2)
\end{pmatrix}.
\end{aligned}
\end{equation}
Note that, for notational simplicity, the superscripts distinguishing down-type and up-type quark Yukawa couplings have been omitted in the above expressions, i.e., $y_i = y_{i}^{(d,u)}$. The determinants of mass squared matrices $M_d M_d^\dagger$ and $M_u M_u^\dagger$ can be uniformly written as:
\begin{equation}
\det(M_d M_d^\dagger, M_u M_u^\dagger) = \frac{|y_2|^2 |y_4|^2}{8}
\left| (v_1^2 v_4 + v_2^2 v_3) y_1 - (v_1 v_3^2 + v_2 v_4^2) y_3 \right|^2.
\end{equation}
To ensure that all quarks have non-zero masses (i.e., no zero-mass eigenvalues) and that the CKM matrix is non-block-diagonal, from the mass squared matrices and determinants, the Yukawa couplings must strictly satisfy the following conditions:
\begin{equation}
\label{quarkckmconditions}
    y_2 \neq 0,\quad  y_4 \neq 0,\quad (v_1^2 v_4 + v_2^2 v_3) y_1 \neq (v_1 v_3^2 + v_2 v_4^2) y_3,
\end{equation}
where $y_1$ and $y_3$ cannot be simultaneously zero, but only one of them is allowed to vanish. 

In Ref.~\cite{Fu:2026phe}, we have shown that, within the framework of spontaneous CP violation, taking the most general complex vacuum C-N-3a as an example, all vacua that can realize spontaneous CP violation are forced by the strong constraints of the $D_5$ symmetry to satisfy the modulus relations $|v_1| = |v_2|$ and $|v_3| = |v_4|$. This in turn leads to a highly symmetric structure of the up- and down-type quark mass-squared matrices in the first two generations, such that they can be simultaneously block-diagonalized by the same unitary transformation $U_0$, ultimately rendering the CKM matrix block-diagonal with strictly vanishing mixing between different blocks, and thus unable to generate non-zero mixing angles and CP violation. The same conclusion applies to the case of real vacua: for the vacuum R-N-2~\cite{Fu:2026agd}, which satisfies $|v_1| = |v_2|$ and $|v_3| = |v_4|$, it also leads to a block-diagonal CKM matrix. Therefore, in order to prevent the most general real vacuum R-N-4a adopted in this work from degenerating into the vacuum R-N-2, and thus to avoid a block-diagonal CKM matrix, the vacuum R-N-4a must satisfy the following conditions:
\begin{equation}
\label{vevsconditions}
v_1 \neq v_2,\qquad v_3 \neq v_4.
\end{equation}
Furthermore, when the vacuum R-N-4 degenerates into one of the vacua R-I-4a, R-I-4b, R-I-4c, or
R-I-4d, it does not lead to a block-diagonal CKM matrix. Therefore, we allow one of the $v_i$ ($i = 1,2,3,4$) in R-N-4 to vanish, while the remaining three moduli are non-zero.

\section{Numerical fitting}
\label{Sec:quarkfit}
Since the mass-squared matrices are structurally complicated and admit no simple closed-form analytic diagonalization, we adopt numerical diagonalization methods. The squared quark masses are obtained from the eigenvalues of the mass-squared matrices, and the CKM matrix is extracted from the diagonalizing matrices. Our goal is to find the global best fit  set of VEVs and quark Yukawa couplings that reproduce the experimental observables, i.e., the quark masses, the CKM matrix elements, and the Jarlskog invariant~\cite{Jarlskog:1985ht},
\begin{equation}
\label{jarlskoq}
J_{\rm CP} = {\rm Im}( V_{ud} V_{cs} V_{us}^* V_{cd}^*).
\end{equation}

\subsection{Standard Model Running Quark Mass}

In new physics beyond the SM, to explain the observed fermion masses, one typically extrapolates them to a common momentum scale $\mu$ far above the QCD scale (about $1\,\mathrm{GeV}$) and even the electroweak scale ($246\,\mathrm{GeV}$) for comparison. Below the electroweak scale, this extrapolation must account for the renormalization group evolution of the mass parameters induced by QCD and QED loop corrections. For heavy quarks ($c,\,b,\,t$), the pole mass $M_q$ is a physical mass (observable), which appears as the pole of the propagator in perturbation theory; for light quarks ($u,\,d,\,s$), due to the non-perturbative nature of the strong interaction at their mass scale, the concept of the pole mass is not adequate (not well-defined). The running mass $m_q(\mu)$ depends on the renormalization scale $\mu$. When taken at $\mu = M_q$, it is denoted as $m_q(M_q)$, which includes QCD and QED loop corrections and can differ significantly from the pole mass $M_q$~\cite{Babu:2009fd}.

For momentum scales above the electroweak symmetry breaking scale, one should evolve the Yukawa couplings of the fermions rather than their masses themselves. At this momentum scale, the running mass is equivalently defined as
\begin{equation}
\label{massltevmu}
m_i(\mu) = y_i(\mu) ~\frac{v}{\sqrt{2}} \ ,
\end{equation}
where $y_i(\mu)$ denotes the running Yukawa coupling of the quark at the reference scale, and $v = 246\ \mathrm{GeV}$ is the VEV of the Higgs doublet at the electroweak scale. Since the VEV $v$ is also a function of momentum (due to the wave function renormalization of the Higgs field), one should in principle use $v(\mu)$ in the above definition to define the running mass. However, this is typically unnecessary, and the present work adopts the definition of Eq.~(\ref{massltevmu}), taking $v = 246\ \mathrm{GeV}$ as a fixed constant and evolving only the Yukawa couplings $y_i(\mu)$ via the renormalization group equations (RGEs); i.e., all running effects are absorbed into the Yukawa couplings, and the running masses are then calculated using the formula defined above. This is because the running of $v(\mu)$ is extremely slow. From the electroweak scale up to $1\ \mathrm{TeV}$ or even to the GUT scale, the variation of $v(\mu)$ is negligible compared with the orders-of-magnitude evolution of the Yukawa couplings $y_i(\mu)$, and the uncertainty thus introduced is far smaller than other theoretical uncertainties.

The $D_5$ 4HDM investigated in this work contains new scalar particles with masses we assume around several $\text{TeV}$. We naturally expect the masses of these new particles are all above $1~\text{TeV}$, thus, their contributions decouple in the range $\mu \leq 1~\text{TeV}$, and the SM renormalization group evolution is therefore a valid approximation up to that scale, and we take the matching scale at $\mu = 1\ \text{TeV}$. Under this premise, Ref.~\cite{Antusch:2025fpm} used the low-energy data from \textbf{2024 PDG}~\cite{ParticleDataGroup:2024cfk} as input, and provided the SM running parameters and their $1\sigma$ highest posterior density (HPD) intervals at various reference scales in the $\overline{\text{MS}}$ renormalization scheme. The quark running Yukawa couplings at the scale $\mu = 1~\text{TeV}$ provided are:
\begin{equation}
\begin{aligned}
&y_u = (6.15\pm 0.14)  \times 10^{-6},\\ 
&y_c = (3.11\pm 0.05) \times 10^{-3},\\  
&y_t = 0.8616\pm 0.0043, \\
&y_d = (1.35\pm 0.02) \times 10^{-5} ,\\
&y_s = (2.68\pm 0.03) \times 10^{-4},\\
&y_b = (1.401\pm 0.009) \times 10^{-2}. 
\end{aligned}
\end{equation}
Based on these parameters with Eq.~(\ref{massltevmu}), we obtain the running quark masses at $\mu = 1~\text{TeV}$ as:
\begin{equation}
\label{runningquark}
\begin{aligned}
m_u &= (0.00107 \pm 0.000024)\ \text{GeV}, \\ 
m_c &= (0.54114 \pm 0.0087)\ \text{GeV},   \\
m_t &= (149.9184 \pm 0.7482)\ \text{GeV},  \\
m_d &= (0.002349 \pm 0.000035)\ \text{GeV}, \\
m_s &= (0.046632 \pm 0.000522)\ \text{GeV},\\
m_b &= (2.43774 \pm 0.01566)\ \text{GeV}.
\end{aligned}
\end{equation}
These parameters serve as the input for our numerical fit.
\subsection{Fit Results of R-N-4a Model}
We perform a numerical fit to the quark mass matrices of the R-N-4a model, i.e., those in Eqs.~(\ref{equ:quark_model_d}) and (\ref{equ:quark_model_u}).
We impose basic constraints on the corresponding parameters through Eqs.~(\ref{quarkckmconditions}) and (\ref{vevsconditions}); moreover, for the real vacuum R-N-4a, the VEVs satisfy $v_1^2+v_2^2+v_3^2+v_4^2 = (246\;\text{GeV})^2$.
These conditions set the initial ranges of the parameters, while the precise ranges are determined by numerical methods.
In the numerical fit, we employed the Differential Evolution algorithm \cite{Storn:1997uea,Price05wholebook,DasSuganthan11,DarkMachinesHighDimensionalSamplingGroup:2021wkt} to scan the parameter space, and obtained the best fit parameters based on a likelihood function constructed from theoretical predictions, observational data, and the corresponding uncertainties.

The explicit form of the likelihood function adopted in this paper 
is given as follows;
\begin{eqnarray}
L(\theta_i) = \prod_{j=1}^{n_{\rm obs}} \frac{1}{\sqrt{2\pi}\,\sigma_j}
\exp\!\left[-\frac{(O_j - f_j(\theta_i))^2}{2\sigma_j^2}\right].
\end{eqnarray}
We here assume Gaussian errors ${\sigma_j}$ for the observables $\{O_j\}$, in thate case, 
since $\ln L(\theta_i) = -\frac{1}{2} \chi^2(\theta_i) + const.$,
maximization of the likelihood function is equivalent to minimizing the $\chi^2$ function, 
\begin{eqnarray}
\chi^2(\theta_i) =\sum_j \chi_j^2(\theta_i), \ \ \  {\rm where}\ \ \ \chi^2_j(\theta_i)= \frac{(O_j - f_j(\theta_i))^2}{\sigma_j^2}.
\end{eqnarray}

The 16 experimental measurements $\{O_j\}$ listed in Table~\ref{quark_measurements_model} ~ \cite{ParticleDataGroup:2024cfk} including the quark mass observables given in Eq.~(\ref{runningquark}) and the CKM matrix elements, are fitted using $f_j$, the quark mass matrices given in Eqs.~(\ref{equ:quark_model_d}), (\ref{equ:quark_model_u}), and the CKM expression in Eqs.~(\ref{equ:CKM})-(\ref{Dmd2}) described by the 12 input (partly complex) parameters $\theta_i=\{v_1,\,v_2,\,v_3,\,v_4,\,y_{1}^d,\,y_2^d,\,y_3^d,\,y_4^d, \,y_1^u,\,y_2^u,\,y_3^u,\,y_4^u\}$. 
Note that $\{y_i^{d,u}\}$ parameters are originally complex but the target function $f_j$ is invariant under a certain symmetry explained in Appendix~\ref{complexphases}, thus, we can remove 5 phases out of 8 phases of them to leave 3 phases. The actual scan is performed fixing all parameters real other than $y_3^d$, $y_3^u$, and $y_4^u$.
Moreover, since $v_{\rm sm}^2=\sum_i^4 v_i^2 = 246 ({\rm GeV})^2$ is also additionally constrained by the other electroweak measurements, we keep this relation always saturated, which reduces one degree of freedom of the input parameters $\theta_i$. Consequently, the effective dimension of the real input parameters becomes 14.
We obtained a set of best fit results, and the corresponding observables are also listed in Table~\ref{quark_measurements_model}.
\medskip

\begin{table}[ht]
    \centering
 \setlength{\tabcolsep}{2.5pt}
\renewcommand{\arraystretch}{1.2}
\begin{tabular}{|c|c|c|c|}
\hline
\multirow{2}{5em}{Observables}&\multicolumn{3}{c|}{R-N-4a Model ($D_5$ 4HDM)}\\
\cline{2-4}
&Input&Best Fit&Pull($\sigma$)\\
\hline
$m_u(\mathrm{GeV})$&0.00107$\pm$0.000024&0.0010700&0.00 \\
$m_c(\mathrm{GeV})$&0.54114$\pm$0.0087&0.541176&0.00 \\
$m_t(\mathrm{GeV})$&149.9184$\pm$0.7482&149.91694&0.00 \\
$m_d(\mathrm{GeV})$&0.002349$\pm$0.000035&0.0023490&0.00 \\
$m_s(\mathrm{GeV})$&0.046632$\pm$0.000522&0.0466369&0.01 \\
$m_b(\mathrm{GeV})$&2.43774$\pm$0.01566&2.437821&0.01 \\
$|V_{ud}|$&0.97367$\pm$0.00032&0.974313&2.01 \\
$|V_{us}|$&0.22431$\pm$0.00085&0.225169&1.01\\
$|V_{ub}|$&0.00382$\pm$0.0002&0.003769&0.26 \\
$|V_{cd}|$&0.221$\pm$0.004&0.2250&1.00 \\
$|V_{cs}|$&0.975$\pm$0.006&0.9735&0.25\\
$|V_{cb}|$&0.0411$\pm$0.0012&0.04182&0.60 \\
$|V_{td}|$&0.0086$\pm$0.0002&0.00857&0.15 \\
$|V_{ts}|$&0.0415$\pm$0.0009&0.04111&0.43 \\
$|V_{tb}|$&1.01$\pm$0.027&0.9991&0.40 \\
$J_{CP}$&$(3.12\pm0.125)\times10^{-5}$&$3.144\times10^{-5}$&0.19 \\
\hline
$\chi^2$&$-$&$-$&6.97 \\
\hline
\end{tabular}
 \caption{The input values and best fit values in the R-N-4a model, together with their pulls at $\mu = 1\,\mathrm{TeV}$, are listed. The observables include six quark masses, nine absolute CKM matrix elements, and the Jarlskog invariant $J_{\mathrm{CP}}$ related to the CKM phase.}
    \label{quark_measurements_model}
\end{table}

The best fit VEVs are:
\begin{equation}
\begin{aligned}
v_1 &= 28.0361\ \text{GeV}, &\quad  v_2 &= 69.8436\ \text{GeV}, \\
v_3 &= 108.4764\ \text{GeV}, & v_4 &= 207.5686 \text{GeV}. 
\end{aligned}
\end{equation}
The best fit quark Yukawa couplings are:
\begin{equation}
\begin{aligned}
y_{1}^d &= 1.56\times10^{-3}, &\quad y_{1}^u &= 4.91\times10^{-2},\\
y_{2}^d &= -5.00\times10^{-5}, & y_{2}^u &= -1.23\times10^{-4},\\
y_{3}^d &= (5.70 - 1.89i)\times10^{-4}, & y_{3}^u &= (9.795 + 0.534i)\times10^{-3},\\
y_{4}^d &= -1.47\times10^{-2}, & y_{4}^u &= (-5.276 - 7.354i)\times10^{-1}.
\end{aligned}
\end{equation}\\

From the above parameter values, we obtain the corresponding quark mass matrices $M_d$ and $M_u$, as follows,

\begin{equation}
\label{downresult2}
M_d = \begin{pmatrix}
(4.372 - 1.450i)\times10^{-2} & 3.096\times10^{-2} & -2.448\times10^{-3} \\
7.712\times10^{-2} & (8.366 - 2.775i)\times10^{-2} & -9.827\times10^{-4} \\
-2.158 & -1.128 & 0
\end{pmatrix}\ \text{GeV}\ ,
\end{equation}\\
\begin{equation}
\label{upresult2}
M_u = \begin{pmatrix}
1.438 + 0.0784i & 2.425 & -2.446\times10^{-3} \\
9.733\times10^{-1} & (7.513 + 0.410i)\times10^{-1} & -6.093\times10^{-3} \\
(-4.047 - 5.641i)\times10^{1}  & (-0.7743 - 1.079i)\times10^{2} & 0
\end{pmatrix}\ \text{GeV}\ .
\end{equation}

\subsection{Analysis of the Fit Results} 
The CP violation originates entirely from the complex phases of the Yukawa couplings: In the down-type quark sector, $\operatorname{Im}(y_{3}^d) = -1.89\times10^{-4}$ gives a small contribution to CP violation. In the Up-type quark sector: $\operatorname{Im}(y_{4}^u) = -7.354\times10^{-1}$ dominates the CP violation in the quark sector, while $\operatorname{Im}(y_{3}^u) = 0.534\times10^{-3}$ provides a small CP-violating contribution. As can be seen from the quark mass matrices in Eqs.~(\ref{equ:quark_model_d}) and (\ref{equ:quark_model_u}), CP violation arises only from the couplings of the $(\phi_3, \phi_4)$ doublets to the up- and down-type quarks. These are the natural sources of CP violation within the $D_5$ symmetry framework. The resulting Jarlskog invariant $J_{\mathrm{CP}} = 3.144\times10^{-5}$ agrees with the experimental value $3.12\times10^{-5}$ within $0.19\sigma$.

The four VEVs exhibit a clear hierarchy:
    \begin{equation}
        v_4 : v_3 : v_2 : v_1 \ \approx \ 7.40 : 3.87 : 2.49 : 1 \ ,
    \end{equation}
i.e., their magnitudes decrease gradually. The down-type and up-type Yukawa couplings (abbreviated as $y_i = y_{i}^{(d,u)}$ ) exhibit the same pronounced hierarchy: $|y_{4}| \gg |y_{1}| \gg |y_{3}|\gg |y_{2}|$, with:
    \begin{equation}
|y_{4}| : |y_{1}| : |y_{3}| : |y_{2}| \approx \ 10^3 : 10^2 : 10 : 1 \ ,
    \end{equation}
 i.e., their magnitudes decrease sharply in this order. 

 In the R-N-4a model, each of the four VEVs and the four independent quark Yukawa coupling coefficients possess a hierarchy after the spontaneous breaking of the $D_5$ symmetry, and they produce the mass hierarchy of the up-type and down-type quarks as well as the hierarchy of the CKM mixing as the consequence of the cross-mixing products in the mass matrices given in Eqs.~(\ref{equ:quark_model_d}) and (\ref{equ:quark_model_u}).

\begin{table}[ht]
    \centering
    \setlength{\tabcolsep}{2.9pt}
    \renewcommand{\arraystretch}{1.3}
    \begin{tabular}{|c|c|}
    \hline
    Physical source & $\chi^2$ contribution \\
    \hline
    Down-type quarks ($m_d, m_s, m_b$)  & 0.00011  \\
    Up-type quarks ($m_u, m_c, m_t$)    & 0.00005  \\
    CKM matrix (9 matrix elements)     & 6.935 \\
    Jarlskog invariant $J_{CP}$        & 0.0378  \\
    \hline
    \textbf{Total $\chi^2$} & \textbf{6.97}   \\
    \hline
    \end{tabular}
    \caption{$\chi^2$ breakdown of the R-N-4a model.}
    \label{tab:chisq_model}
\end{table}
 
The $\chi^2$ decomposition of the R-N-4a model is shown in Table~\ref{tab:chisq_model}, with a total $\chi^2 = 6.97$, indicating an excellent fit. The CKM part contributes almost all $\chi^2$, with the pull of $|V_{ud}|$ around $2\sigma$ being the largest single source; the pulls of $|V_{us}|$ and $|V_{cd}|$ are about $1\sigma$. The fit to the quark masses is nearly perfect.
It should be noted that the model contains 14 real free parameters: 3 real VEVs $v_1, v_2, v_3$ (with $v_4^2 = v^2 -\sum_i^3 v_i^2)$; 3 complex Yukawa couplings $y_{3}^d, y_{3}^u, y_{4}^u$ (each contributes 2 real parameters); and 5 real Yukawa couplings $y_{1}^{d}, y_{2}^{d}, y_{4}^{d}, y_{1}^{u}, y_{2}^{u}$, where we can show these 5 yukawa parameters can be fixed real without affecting the observables. Therefore, these 14 real parameters are used to fit all 16 experimental observables in the quark sector (6 quark masses, 9 CKM matrix elements, and 1 Jarlskog invariant).

\section{Conclusion and Outlook}
\label{Sec:Conclusion}
In this paper, we have investigated the quark sector of a four-Higgs-doublet model with an exact $D_5$ symmetry. We adopt the most general real vacuum, namely the R‑N‑4a model, and introduce explicit CP violation through some of the complex phases in the Yukawa couplings. Using a Differential Evolution algorithm, we perform a numerical fit of the quark masses and the CKM matrix (16 observables in total) with 14 free real parameters, and obtain an excellent total chi-squared value $\chi^2 = 6.97$. The quark masses are fitted with high precision, with all deviations below $0.01\sigma$; the nine moduli of the CKM matrix elements are well reproduced, with only $|V_{ud}|$ showing a slightly larger deviation of $2\sigma$, while all the others are within about $1\sigma$. The CP violation originates from the complex phases of the three Yukawa couplings, for example we can take $y_{3}^{d}$, $y_{3}^{u}$, $y_{4}^{u}$, and the resulting Jarlskog invariant $J_{\mathrm{CP}} = 3.144\times10^{-5}$ agrees with the experimental value within $0.19\sigma$. Therefore, its intrinsic new sources of CP violation offer a novel path for understanding the observed CP violation in the quark sector. Crucially, the intrinsic hierarchies of the vacuum expectation values and the Yukawa coupling elements, allowed by the $D_5$ symmetry and its spontaneous breaking, can enter the quark mass matrices through cross-mixing products and thereby naturally give rise to the observed quark mass and mixing hierarchy.

Although a detailed phenomenological analysis is beyond the scope of this work, we briefly comment on the flavor-changing couplings. The $D_5$ flavor symmetry and the adopted vacuum structure are designed to address the quark flavor puzzle and do not forbid tree-level FCNCs/FCCCs in the neutral/charged scalar sector. In the general parameter space, the neutral Higgs bosons couple to different quark flavors, generating tree-level contributions to FCNC meson mixing, while the charged Higgs bosons can induce FCCC processes at tree level. We note that this model itself does not forbid FCNC couplings even for the 125 GeV Higgs. These couplings are strongly constrained by flavor observables while top-charm FCNC couplings are relatively weakly constrained, and the consequent $t\to ch$ decays are expected to be important signatures at the LHC.
A special choice of parameters may realize the Higgs alignment limit, where the 125 GeV Higgs couplings become flavor-diagonal so that it does not mediate tree-level FCNCs. Even in that case, however, the heavy neutral Higgs bosons generically still have flavor-changing couplings and can mediate tree-level FCNCs. The most stringent constraints arise from $K^0-\bar{K}^0$ mixing and $B_s\to\mu^+\mu^-$, which typically require the new scalar masses to be above the $\mathcal{O}(1)$ TeV scale or even beyond unless the relevant Yukawa couplings are sufficiently suppressed. Partial cancellations among the extra parameters in the 4HDM can relax these bounds. Future experiments such as Belle II will provide stronger bounds on these couplings. A dedicated analysis will be presented in a forthcoming publication.

\section*{Acknowledgments}
This work is supported by the Fundamental Research Funds for the Central Universities, the One Hundred Talent Program of Sun Yat-sen University, China, and the Guangdong Natural Science Foundation (Project No. 2026A1515012641).

\appendix

\section{$D_5$-symmetric 4HDM potential}

The most general scalar potential of $D_5$-symmetric 4HDM is given~\cite{Fu:2026agd} as follows:
\begin{equation}
\begin{aligned}
\label{V}
V=&-\mu^2_{1}(\phi^\dagger_1\phi_1+\phi^\dagger_2\phi_2)-\mu^2_{2}(\phi^\dagger_3\phi_3+\phi^\dagger_4\phi_4)+\lambda_1\left[(\phi^\dagger_1\phi_1)^2+(\phi^\dagger_2\phi_2)^2\right]\\&+
\lambda_2\left[(\phi^\dagger_3\phi_3)^2+(\phi^\dagger_4\phi_4)^2\right]+\lambda_3(\phi^\dagger_1\phi_1)(\phi^\dagger_2 \phi_2)+\lambda_4(\phi^\dagger_3\phi_3)(\phi^\dagger_4 \phi_4)\\
&+\lambda_5\left[(\phi^\dagger_1\phi_1)(\phi^\dagger_3\phi_3)+(\phi^\dagger_2\phi_2)(\phi^\dagger_4\phi_4)\right]+\lambda_6\left[(\phi^\dagger_2\phi_2)(\phi^\dagger_3\phi_3)+(\phi^\dagger_1\phi_1)(\phi^\dagger_4\phi_4)\right]\\
&+\lambda_{7}(\phi^\dagger_1\phi_2)(\phi^\dagger_2\phi_1)
+\lambda_{8}(\phi^\dagger_3\phi_4)(\phi^\dagger_4\phi_3)+\lambda_{9}\left[(\phi^\dagger_1\phi_3)(\phi^\dagger_3\phi_1)+(\phi^\dagger_2
\phi_4)(\phi^\dagger_4\phi_2)\right]\\
&+\lambda_{10}\left[(\phi^\dagger_1\phi_4)(\phi^\dagger_4\phi_1)+(\phi^\dagger_2\phi_3)(\phi^\dagger_3\phi_2)\right] +\left\{ 
\lambda_{11}(\phi^\dagger_1\phi_3)(\phi^\dagger_2\phi_4)
+\lambda_{12}(\phi^\dagger_2\phi_3)(\phi^\dagger_1\phi_4)
\right.\\
&\left.+\lambda_{13}\left[(\phi^\dagger_2\phi_1)(\phi^\dagger_2\phi_3)+(\phi^\dagger_1\phi_2)(\phi^\dagger_1\phi_4)\right]
+\lambda_{14}\left[(\phi^\dagger_1\phi_3)(\phi^\dagger_4\phi_3)+(\phi^\dagger_2\phi_4)(\phi^\dagger_3\phi_4)\right]
+\text{h.c.} \right\}.
\end{aligned}
\end{equation}

\section{Tensor product of $D_5$}

In $D_5$, the Clebsch-Gordan coefficients for all tensor products are given as follows. 

A singlet $w$ and a singlet $z$ give
\begin{equation}
\label{D511}
\begin{aligned}
&(w)_{\mathbf{1}}\otimes  (z)_{\mathbf{1}}=(w)_{\mathbf{1}'}\otimes (z)_{\mathbf{1}'} =(wz)_{\mathbf{1}}, \qquad (w)_{\mathbf{1}}\otimes  (z)_{\mathbf{1}'}=(wz)_{\mathbf{1}'}.
\end{aligned}
\end{equation}

A singlet $w$ and a doublet $x=(x_1, x_2)^\intercal$, where $\bar{\mathbf{2}}=\mathbf{2}$ and $\mathbf{2}'$, give
\begin{equation}
\label{D5C12}
\begin{aligned}
&(w)_{\mathbf{1}}\otimes 
\begin{pmatrix}
x_1 \\
x_2 
\end{pmatrix}_{\bar{\mathbf{2}} }
=\begin{pmatrix}
wx_1 \\
wx_2 
\end{pmatrix}_{\bar{\mathbf{2}}},\qquad
(w)_{\mathbf{1}'}\otimes 
\begin{pmatrix}
x_1 \\
x_2 
\end{pmatrix}_{\bar{\mathbf{2}}}
=\begin{pmatrix}
wx_1 \\
-wx_2 
\end{pmatrix}_{\bar{\mathbf{2}}}.
\end{aligned}
\end{equation}

Two doublets, $x=(x_1 , x_2)^\intercal$ and $ y=(y_1 , y_2)^\intercal$, give
\begin{equation}
\label{D5C22}
\begin{aligned}
&\begin{pmatrix}
x_1 \\
x_2 
\end{pmatrix}_{\mathbf{2}}\otimes 
\begin{pmatrix}
y_1 \\
y_2 
\end{pmatrix}_{\mathbf{2}}
=(x_1y_2+x_2y_1)_{\mathbf{1}}
\oplus (x_1y_2-x_2y_1)_{\mathbf{1}'}
\oplus 
\begin{pmatrix}
x_1y_1 \\
x_2y_2 
\end{pmatrix}_{\mathbf{2}'}, \\&
\begin{pmatrix}
x_1 \\
x_2 
\end{pmatrix}_{\mathbf{2}'}\otimes 
\begin{pmatrix}
y_1 \\
y_2 
\end{pmatrix}_{\mathbf{2}'}
=(x_1y_2+x_2y_1)_{\mathbf{1}}
\oplus (x_1y_2-x_2y_1)_{\mathbf{1}'}
\oplus 
\begin{pmatrix}
x_2y_2 \\
x_1y_1 
\end{pmatrix}_{\mathbf{2}}, \\&
\begin{pmatrix}
x_1 \\
x_2 
\end{pmatrix}_{\mathbf{2}}\otimes 
\begin{pmatrix}
y_1 \\
y_2 
\end{pmatrix}_{\mathbf{2}'}
=\begin{pmatrix}
x_2y_1 \\
x_1y_2
\end{pmatrix}_{\mathbf{2}} 
\oplus 
\begin{pmatrix}
x_2y_2 \\
x_1y_1
\end{pmatrix}_{\mathbf{2}'}. 
\end{aligned}
\end{equation}

\section{Unitary mixing matrices}

By numerically diagonalizing the quark mass matrices $M_d$ and $M_u$, the left-handed and right-handed unitary mixing matrices for the down-type and up-type quarks, namely $U_{dL}, U_{dR}, U_{uL}, U_{uR}$, can be obtained. The results are shown in
Equations~(\ref{quarkt2-1}) - (\ref{quarkt2-4}):

\begin{equation}
\label{quarkt2-1}
U_{d_L} = \begin{pmatrix}
0.977   &       0.211      &    0.0224  \\
-0.0533+0.205i&  0.242-0.946i & 0.0440+0.00521i \\
 0.0179+0.0139i& 0.0202-0.0392i& -0.971-0.235i
\end{pmatrix},
\end{equation}
\begin{equation}
\label{quarkt2-2}
U_{d_R} = \begin{pmatrix}
0.0116+0.00262i & 0.339-0.317i&  0.861+0.208i \\
-0.0222-0.00504i& -0.647+0.604i&   0.451+0.109i \\
-0.996-0.0856i & 0.0162-0.0199i&  -0.0000402-0.00000210i
\end{pmatrix},
\end{equation}
\begin{equation}
\label{quarkt2-3}
U_{u_L} = \begin{pmatrix}
0.952 & 0.307 &  0.0188\\
-0.274+0.139i & 0.849-0.429i& 0.00745 +0.000146i \\
0.00854 +0.00134i & 0.00981+0.00771i&-0.593-0.805i
\end{pmatrix},
\end{equation}
\begin{equation}
\label{quarkt2-4}
U_{u_R} = \begin{pmatrix}
 0.00665+0.00827i&  0.805-0.371i& 0.463-0.00598i \\
-0.00347-0.00432i& -0.421+0.194i&0.886-0.0114i \\
0.615+0.789& -0.0109+0.00484i&-0.000000609-0.00000000594i
\end{pmatrix}.
\end{equation} 
\

\section{Complex phases of the yukawa parameters}
\label{complexphases}
For the two mass matrices of down-type and up-type quarks we can consider how many degrees of freedom (phases) leave the six mass eigenvalues and the CKM matrix invariant. We see in our model, 5 phases can be rotated away and 3 would be physical, which we selected $y_{3}^{d}$, $y_{3}^{u}$, $y_{4}^{u}$ in this paper.

\begin{itemize}
    \item[1.] The quark mass terms are given by
\[
\bar{u}_L^\prime M_u u_R^\prime + \bar{d}_L^\prime M_d d_R^\prime + \text{h.c.}
\]
where \(M_u\) and \(M_d\) are \(3\times 3\) complex matrices, and left-handed doublet in flavor basis is given by $q_L = (u_L^\prime, d_L^\prime)^T$. 
\item[2.] 
One can diagonalize them with unitary transformations:
\[
U_u^\dagger M_u V_u = \operatorname{diag}(m_u, m_c, m_t), \qquad
U_d^\dagger M_d V_d = \operatorname{diag}(m_d, m_s, m_b), 
\]
where each mass eigenvalue real and positive, and the CKM matrix is described as \(V_{\text{CKM}} = U_u^\dagger U_d\).

\item[3.]
We have 3 left-handed and 6 right-handed quark fields.  
Each can be rotated by an independent phase:
\[
q_{Li} \to e^{i\theta_{Li}} q_{Li},
\quad u_{Rj}^\prime \to e^{i\theta_{Rj}^u} u_{Rj}^\prime,
\quad d_{Rj}^\prime \to e^{i\theta_{Rj}^d} d_{Rj}^\prime
\]
Initially, that gives 9 phase parameters. Under these rephasings we can consider the transformed mass matrices:
\[
(M_u)_{ij} \to e^{-i\theta_{Li}} (M_u)_{ij} e^{i\theta_{Rj}^u},\quad
(M_d)_{ij} \to e^{-i\theta_{Li}}  (M_d)_{ij}
e^{i\theta_{Rj}^d}.\]
\item[4.]
Starting with these mass matrices, choosing 
\[
U_u^\dagger \to U_u^\dagger e^{i\theta_{Li}},
V_u \to e^{-i\theta_{Ri}^u} V_u, 
U_d^\dagger \to U_d^\dagger e^{i\theta_{Li}},
V_d \to e^{-i\theta_{Ri}^d} V_d
\] will diagonalize the mass matrices as before and keep the CKM matrix $V_{CKM} = U_u^\dagger U_d$ invariant.

\item[5.]
Using these 9 phases, from 
Eqs.~(\ref{equ:quark_model_d}) and Eqs.~(\ref{equ:quark_model_u}), one can remove 5 phases to leave only 3 parameters complex, for example, $y_3^d, y_3^u$ and $y_4^u$.
\end{itemize}

\end{document}